\documentclass[aps,prd,preprint,twocolumns,showpacs]{revtex4}
\usepackage{epsfig}
\usepackage{amsmath}
\usepackage[hyperref]{hyperref}

\usepackage{color}

\begin{document}

\title{On the mistake in the implementation of the minimal model of the dual parameterization
and resulting inability to describe the high-energy DVCS data}

\author{V. Guzey}
\email{vguzey@jlab.org}
\affiliation{Theory Center, Jefferson Laboratory, Newport News, VA 23606, USA}

\author{T. Teckentrup}
\email{tobias.teckentrup@tp2.rub.de}
\affiliation{Institut f{\"u}r Theoretische Physik II, Ruhr-Universit{\"a}t Bochum,
  44780 Bochum, Germany}

\preprint{JLAB-THY-08-897}

\begin{abstract}

We correct the mistaken claim made in~\cite{Guzey:2005ec,Guzey:2006xi} that the minimal model of the dual parameterization of nucleon generalized parton distributions (GPDs)
gives a good, essentially model-independent description of 
high-energy data on deeply virtual Compton scattering (DVCS).
In the implementation of the dual parameterization in~\cite{Guzey:2005ec,Guzey:2006xi},
 the numerical prefactor of two in front of the DVCS amplitude was missing. 
We show that the corrected minimal model of the dual parameterization significantly 
overestimates the HERA data (H1 and ZEUS) on the DVCS cross section.

\end{abstract}

\maketitle


The dual parameterization of nucleon generalized parton distributions
(GPDs) represents the nucleon GPDs as an infinite series of light-cone distribution
amplitudes in the $t$-channel, which motivates the name of the parameterization~\cite{Polyakov:2002wz}.
The dual parameterization provides a convenient and flexible model for nucleon
GPDs. In particular, the resulting GPDs 
 obey polynomiality and evolve in $Q^2$ according
to the usual forward DGLAP evolution equation (to the leading order in the strong coupling
constant).
In addition, 
the dual parameterization has a well-controlled small-$\xi$ behavior, which
makes the parameterization especially useful at large energies.

Within the framework of the dual parameterization, the following
theoretical questions have recently been addressed: the amount of information on GPDs
that can be extracted from the amplitudes of exclusive 
processes (essentially, the inversion problem, i.e.~the representation 
of a certain function of GPDs in terms of the measurable deeply virtual Compton scattering 
(DVCS) amplitude, 
was solved)~\cite{Polyakov:2007rv,Polyakov:2007rw}; 
the formulation of the dual parameterization and the solution of the inversion 
problem up to the twist-three accuracy~\cite{Moiseeva:2008qd};
the small-$\xi$ limit and the inversion of the dual parameterization in this limit~\cite{SemenovTianShansky:2008mp}.
Phenomenological applications of the dual parameterization include 
applications to the Jefferson Lab data (unpolarized DVCS cross section and the beam helicity
cross section difference)~\cite{Polyakov:2008xm} and to
the high-energy HERA (ZEUS and H1) and HERMES data~\cite{Guzey:2005ec,Guzey:2006xi}.
Unfortunately, the particular implementation of the dual parameterization 
(the so-called minimal model of the dual parameterization) in~\cite{Guzey:2005ec,Guzey:2006xi} used the non-standard
normalization of GPDs, which resulted in a missing factor of two in 
front of the DVCS amplitude. With the factor of two included, the minimal model of~\cite{Guzey:2005ec,Guzey:2006xi} fails to describe the high-energy data on the 
unpolarized DVCS cross section measured by ZEUS and H1 and the beam-spin DVCS asymmetry
measured by HERMES.
It is the purpose of this note to correct the mistaken
claim made in~\cite{Guzey:2005ec,Guzey:2006xi} that the minimal model of the dual parameterization 
gives a good, essentially model-independent description of 
high-energy data on DVCS.

Below we recapitulate key expressions of the dual representation, using the standard
normalization (forward limit) of GPDs. Below we shall consider the GPD $H$; the corresponding
expressions for the GPD $E$ are obtained in a similar way. 
In the dual parameterization, the singlet quark
GPD,
$H^q_{\rm singlet}(x,\xi,t) \equiv H^q(x,\xi,t) - H^q(-x,\xi,t)$,
is given by the following expression,
\begin{equation}
H^q_{\rm singlet}
(x,\xi,t,Q^2)=2 \sum_{\substack{n=1\\  {\rm odd}}}^{\infty}
\sum_{\substack{l=0\\  {\rm even}}}^{n+1} B_{nl}^q(t,Q^2) \Theta(\xi-|x|) \left(1-\frac{x^2}{\xi^2} \right) C_n^{3/2}\left(\frac{x}{\xi}\right) P_l\left(\frac{1}{\xi}\right) \,,
\label{eq:H}
\end{equation}
where $q$ is the quark flavor; $B_{nl}^q(t,Q^2)$ are certain functions with the known
$Q^2$ evolution (see below); $C_n^{3/2}$ are Gegenbauer polynomials and
$P_l$ are Legendre polynomials. Note that we introduced the overall factor of two in
Eq.~(\ref{eq:H}) compared to Eq.~(1) of~\cite{Guzey:2006xi}. 

The series in Eq.~(\ref{eq:H}) can be summed
by introducing a set of generating functions $Q_k^q$~\cite{Polyakov:2002wz},
\begin{equation}
B_{n n+1-k}^q(t,Q^2)=\int^1_0 dx\, x^n\, Q_k^q(x,t,Q^2) \,. 
\label{eq:Q}
\end{equation}
With help of the functions $Q_k^q$, the expression for the 
GPD $H^q_{\rm singlet}$
can be written in the following form (for brevity, we shall omit the dependence on the virtuality $Q^2$):
\begin{align}
H^q_{\rm singlet}&(x,\xi,t)=2 \sum_{\substack{n=0\\  {\rm even}}}^{\infty}\frac{\xi^k}{2}\left(H^{q\,(k)}(x,\xi,t)-H^{q\,(k)}(-x,\xi,t)\right)
\nonumber\\
&+2\left(1-\frac{x^2}{\xi^2}\right) \Theta(\xi-|x|) \sum_{\substack{l=1\\  {\rm odd}}}^{k-3} C_{k-l-2}^{3/2}\left(\frac{x}{\xi}\right) P_l\left(\frac{1}{\xi}\right) \int^1_0 dy\, y^{k-l-2}\,Q_k^q(y,t) \,,
\label{eq:resumH}
\end{align} 
where
\begin{align}
H^{q\,(k)}
(x,\xi,t)&=\Theta(x-\xi) \frac{1}{\pi}\int^1_{y_0}\frac{dy}{y} \left(1-y\frac{\partial}{\partial y}\right) Q_k^q(y,t) \int^{s_2}_{s_1} ds \frac{x_s^{1-k}}{\sqrt{2x_s-x_s^2-\xi^2}} 
 \nonumber\\
&+\Theta(\xi-x) \Bigg\{\frac{1}{\pi}\int^1_{0}\frac{dy}{y} \left(1-y\frac{\partial}{\partial y}\right) Q_k^q(y,t) \int^{s_3}_{s_1} ds \frac{x_s^{1-k}}{\sqrt{2x_s-x_s^2-\xi^2}} \nonumber\\
&-\lim_{y\to 0} Q_k^q(y) \int^{s_3}_{s_1} ds \frac{x_s^{1-k}}{\sqrt{2x_s-x_s^2-\xi^2}}
\Bigg\} \,,
\label{eq:Hk}
\end{align}
and $x_s=2 (x-s \xi)/((1-s^2)y)$. The limits of integration in Eq.~(\ref{eq:Hk}) are:
\begin{eqnarray}
y_0&=&\frac{1}{\xi^2}\left[x(1-\sqrt{1-\xi^2})+\sqrt{(x^2-\xi^2)(2(1-\sqrt{1-\xi^2})-\xi^2)}
 \right] \,, \nonumber\\
s_1&=&\frac{1}{y \xi}\left[1-\sqrt{1-\xi^2}-\sqrt{2(1-xy)(1-\sqrt{1-\xi^2})-\xi^2(1-y^2)}
 \right] \,, \nonumber\\
s_2&=&\frac{1}{y \xi}\left[1-\sqrt{1-\xi^2}+\sqrt{2(1-xy)(1-\sqrt{1-\xi^2})-\xi^2(1-y^2)}
 \right] \,, \nonumber\\
s_3&=&\frac{1}{y \xi}\left[1+\sqrt{1-\xi^2}-\sqrt{2(1-xy)(1+\sqrt{1-\xi^2})-\xi^2(1-y^2)}
 \right] \,.
\label{eq:limits}
\end{eqnarray}
The prefactor $\xi^k$ in Eq.~(\ref{eq:resumH}) indicates that in the 
small-$\xi$ limit, only a small finite number of functions
$H^{q\,(k)}$
(generating functions $Q_k^q$) contributes to the GPD 
$H^q_{\rm singlet}$.
The minimal model of~\cite{Guzey:2005ec,Guzey:2006xi} retains only $Q_0^q$ and $Q_2^q$; 
the analysis of~\cite{Polyakov:2008xm} uses only $Q_0^q$.

One should emphasize that the generating function
$Q_0^q(x,t)$ is expressed in terms of the $t$-dependent quark parton distribution function (PDF)
$q(x,t)$ ($q(x,t)$ reduces to the usual quark PDFs of flavor $q$ in the $t=0$ limit)~\cite{Polyakov:2002wz},
\begin{equation}
Q_0^q(x,t)=q(x)+\bar{q}(x)-\frac{x}{2}\int^1_{x} \frac{dz}{z^2} \left(q(z)+\bar{q}(z)\right) \,.
\label{eq:Q0}
\end{equation}

Note again that we introduced an overall factor of two in Eq.~(\ref{eq:resumH}) compared to
Eq.~(7) of~\cite{Guzey:2006xi}. This is needed in order to provide the properly normalized 
forward limit of the singlet GPD 
$H^q_{\rm singlet}$.
Indeed, in the $\xi=t=0$ limit, only
the generating function $Q_0^q$ contributes to Eq.~(\ref{eq:resumH}), and using
Eqs.~(\ref{eq:resumH})-(\ref{eq:Q0}),
 one obtains the conventional expression (normalization) for the forward limit 
of the singlet GPD $H^q_{\rm singlet}$
(we assume that $x \geq 0$):
\begin{equation}
H^q_{\rm singlet}(x,0,0)=H^{q\,(0)}(x,0,0)=q(x)+\bar{q}(x) \,.
\label{eq:fl}
\end{equation}

In Refs.~\cite{Guzey:2005ec,Guzey:2006xi}, the factor of two in Eqs.~(\ref{eq:H}) and
(\ref{eq:resumH}) was missing and, as a result, the singlet quark GPD had the unconventional
forward limit (normalization), $H^q(x,0,0)=1/2 (q(x)+\bar{q}(x))$.
In order to evaluate DVCS observables, Refs.~\cite{Guzey:2005ec,Guzey:2006xi} used the 
standard expressions~\cite{Belitsky:2001ns} that implicitly assumed the conventional normalization of
$H^q_{\rm singlet}$ given by Eq.~(\ref{eq:fl}).
Therefore, the predicted values for the DVCS cross section and
the DVCS asymmetries were underestimated by the factors of four and two, respectively.
(Note that the comparison to the double distribution (DD) model and Figs.~1, 2 and 3 
in~\cite{Guzey:2006xi} are correct since the used DD model also had 
the unconventional forward limit.)  
Below we shall give just two examples.

The DVCS amplitude at the photon level
(the Compton form factor, CFF) reads
\begin{equation}
{\cal H}(\xi,t)=\sum_q e_q^2 \int^1_0 dx \, 
H^q_{\rm singlet}(x,\xi,t)
\left(\frac{1}{x-\xi+i \epsilon}+\frac{1}{x+\xi-i \epsilon} \right) \,,
\label{eq:CFF}
\end{equation}
where $e_q$ is the quark electric charge. Using the generating functions $Q_k^q$,
the CFF
can be written in the following convenient form,
\begin{equation}
{\cal H}(\xi,t)=-2 \sum_q e_q^2 \int^1_0 \frac{dx}{x} \sum_{k=0}^{\infty} x^k Q_k^q(x,t)
\left(\frac{1}{\sqrt{1-\frac{2 x}{\xi}+x^2}}+\frac{1}{\sqrt{1+\frac{2 x}{\xi}+x^2}}-2 \delta_{k0} \right) \,.
\label{eq:CFF2}
\end{equation}
Again, note the factor of two in Eq.~(\ref{eq:CFF2}), which is absent in Eq.~(40) 
in~\cite{Guzey:2006xi}. 
Note the implementation of the dual parameterization in~\cite{Polyakov:2008xm} does 
include the factor of two in the expression for ${\cal H}$ and, hence, the predictions
of~\cite{Polyakov:2008xm} do not suffer from the missing factor of two that we
discuss in this note.

At high-energies (small-$\xi$), it is a good approximation to neglect the real part
of the DVCS amplitude and to keep only the contribution of the generating function
$Q_0^q$,
\begin{equation}
\Im m {\cal H}(\xi,t) \approx -2 \sum_q e_q^2 \int^1_{\frac{1-\sqrt{1-\xi^2}}{\xi}} \frac{dx}{x} \,Q_0^q(x,t)
\frac{1}{\sqrt{\frac{2 x}{\xi}-x^2-1}} \,.
\label{eq:ImCFF}
\end{equation}
Therefore, neglecting the suppressed contribution of all GPDs but $H$ to the DVCS amplitude,
we obtain the following simple (approximate) expressions for the unpolarized 
integrated and differential DVCS cross sections, respectively,
\begin{eqnarray}
\sigma_{\rm DVCS}(x_B,Q^2) & \approx & \frac{\pi \alpha_{\rm em} x_B^2}{Q^4}
\int^{t_{\rm min}}_{-1} dt \left(\Im m {\cal H}(\xi,t,Q^2)\right)^2 \,,
\nonumber\\
\frac{d\sigma_{\rm DVCS}(x_B,Q^2)}{dt} & \approx & \frac{\pi \alpha_{\rm em} x_B^2}{Q^4}
 \left(\Im m {\cal H}(\xi,t,Q^2)\right)^2 \,,
\label{eq:sigma}
\end{eqnarray} 
where $\alpha_{\rm em}$ is the fine-structure constant;
Bjorken $x_B= 2\,\xi/(1+\xi)$; 
$t_{\rm min} \approx 0$.
 Note also that we have restored the
$Q^2$-dependence of the CFF ${\cal H}$.

While the present model is simpler than the minimal model 
used in~\cite{Guzey:2005ec,Guzey:2006xi},
the two models are numerically very similar
for the kinematics and observables considered in this note. Therefore, 
we shall also call our model the minimal model; the 
conclusions drawn using the present model also apply to the results 
of~\cite{Guzey:2005ec,Guzey:2006xi}.

The dual parameterization does not model the $t$-dependence of the GPDs, which has to 
be specified 
separately. In this note, we use three models of the $t$-dependence: the exponential and Regge-motivated models used in~\cite{Guzey:2006xi} and the Regge-motivated model of~\cite{Diehl:2007zu}.
These models of the $t$-dependence are applied at the initial
QCD evolution scale, $Q_0^2=1$ GeV$^2$ in our case. 
For unpolarized quark and gluon PDFs, we
use the leading order (LO) CTEQ5L fit~\cite{Lai:1999wy}. 
At each given value of $t$, we use
the usual LO DGLAP evolution in order to obtain the $t$-dependent quark PDF
$q(x,t)$ at the desired scale $Q^2$. Then, using Eqs.~(\ref{eq:Q0}),  
(\ref{eq:ImCFF}) and (\ref{eq:sigma}), we obtain the DVCS amplitude and the DVCS
cross section in the desired kinematics.

\begin{figure}[h]
\begin{center}
\epsfig{file=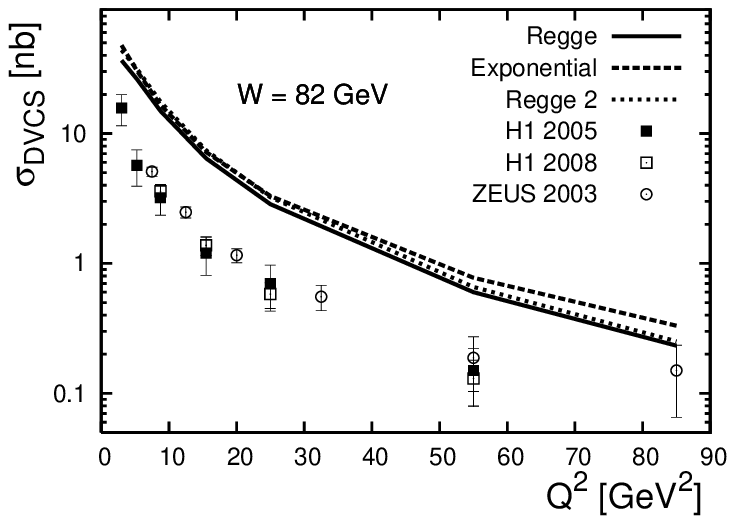,scale=1.2}
\epsfig{file=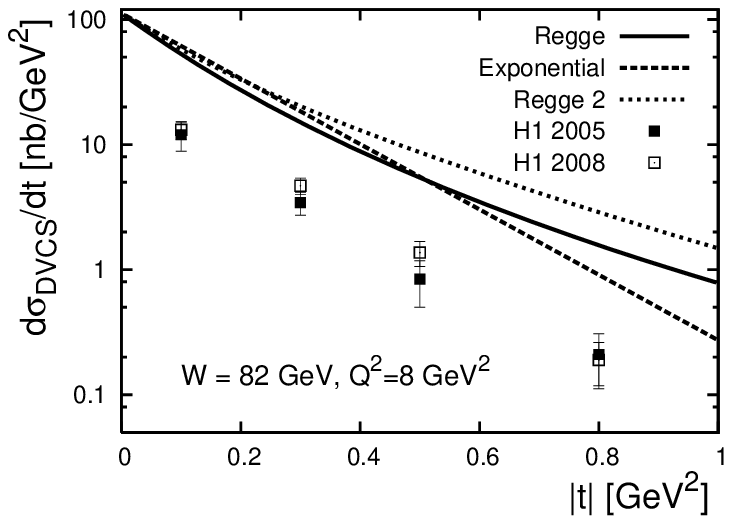,scale=1.2}
\end{center}
\caption{A comparison of the corrected predictions of the dual parameterization 
to the H1~\cite{Aktas:2005ty,aaron:2007cz} and ZEUS~\cite{Chekanov:2003ya}
 DVCS data. The experimental statistical and systematic errors are added in 
quadrature.
The curves correspond to the three used models of the $t$-dependence of GPDs:
the nonfactorized Regge-motivated model of~\cite{Guzey:2006xi} (labeled ''Regge''),
the factorized exponential model of~\cite{Guzey:2006xi} (labeled ''Exponential'')
and the nonfactorized Regge-motivated model of~\cite{Diehl:2007zu}
(labeled ''Regge~2'').
}
\label{fig:dvcs_comparison}
\end{figure}

Figure~\ref{fig:dvcs_comparison} compares predictions of the minimal model of the dual parameterization to the H1~\cite{Aktas:2005ty,aaron:2007cz} and ZEUS data~\cite{Chekanov:2003ya} on the DVCS cross section.
The curves correspond to the three used models of the $t$-dependence:
the nonfactorized Regge-motivated model of~\cite{Guzey:2006xi} (labeled ''Regge''),
the factorized exponential model of~\cite{Guzey:2006xi} (labeled ''Exponential'')
and the nonfactorized Regge-motivated model of~\cite{Diehl:2007zu}
(labeled ''Regge~2''). The message of Fig.~\ref{fig:dvcs_comparison} is clear: once
the missing factor of two is restored, the minimal model of the dual parameterization~\cite{Guzey:2006xi} significantly (by the factor of four)
oversestimates the normalization of the data. Therefore, the claim 
that the minimal model of the dual parameterization 
gives a good, essentially model-independent description of 
high-energy data on DVCS is false.

Similarly, the corrected minimal model of the dual parameterization overestimates
the beam-spin DVCS asymmetry measured by HERMES
by approximately a factor of 1.5
(the increase of the numerator by the factor of two is somewhat offset by the 
increase of the interference term by the factor of two and the DVCS amplitude squared term
by the factor of four), see~\cite{Guzey:2006xi} for references. 
While our minimal model does not describe high-energy DVCS data,
the ''zero step'' model of the dual parameterization 
provides a reasonable description of the data at  lower energies in the Jefferson Lab 
kinematics~\cite{Polyakov:2008xm}.

The minimal model of the dual parameterization at the leading order does not provide any freedom
to adjust the normalization of the DVCS amplitude since it is expressed in terms of the
$t$-dependent quark PDF $q(x,t)$, whose $t$-dependence is constrained essentially by the
H1 DVCS data~\cite{Aktas:2005ty,aaron:2007cz}. Our negative result means that if one intends to
build a successful model for GPDs at large energies, one needs to modify our minimal model, 
e.g.~by keeping more generating functions in Eq.~(\ref{eq:resumH}) or by 
performing the analysis at the next-to-leading order accuracy~\cite{Kumericki:2007sa}.

In summary, a particular implementation of the dual parameterization of nucleon
 GPDs~\cite{Polyakov:2002wz},
the minimal model of the dual parameterization~\cite{Guzey:2005ec,Guzey:2006xi}, 
fails to describe high-energy DVCS data. 
However, this does not mean that the dual parameterization should be discarded: 
different implementations of the dual parameterization in a different kinematics, see
e.g.~\cite{Polyakov:2008xm}, and extended to the next-to-leading order accuracy, see e.g.~\cite{Kumericki:2007sa}, do provide a good description of DVCS observables.

\acknowledgments

We would like to thank D.~Mueller for pointing out the missing factor of two in~\cite{Guzey:2005ec,Guzey:2006xi}
and M.V.~Polyakov and K.~Semenov-Tyan-Shansky for explaining the origin of the missing factor. 
We also would like to thank M.~Diehl for the discussion of the $t$-dependence of the dual parameterization and
for pointing out Ref.~\cite{Diehl:2007zu}.

T.~Teckentrup is supported by the Cusanuswerk.

Authored by Jefferson Science Associates, LLC under U.S. DOE Contract No. DE-AC05-06OR23177. The U.S. Government retains a non-exclusive, paid-up, irrevocable, world-wide license to publish or reproduce this manuscript for U.S. Government purposes.

\end{document}